\documentclass[prl,twocolumn,preprintnumbers,groupedaddress,nofootinbib,floatfix]{revtex4-1}
\usepackage{graphicx}
\usepackage{dcolumn}
\usepackage{bm,amsfonts,amsthm,amsmath,amssymb}
\usepackage[]{epsfig,graphicx}
\usepackage{comment}
\usepackage[T1]{fontenc}
\usepackage{lipsum}
\usepackage[utf8]{inputenc}
\usepackage{ulem}
\usepackage{multirow}
\usepackage{color}
\usepackage{lastpage}
\usepackage[sort&compress]{natbib}
\usepackage{enumerate}
\usepackage{subfigure}
\usepackage{wasysym}
\usepackage{hyperref}
\usepackage{relsize}
\usepackage{physics}
\usepackage{mathrsfs}
\usepackage{tensor}

\hypersetup{
    bookmarks=true,         
    unicode=false,          
    pdftoolbar=true,        
    pdfmenubar=true,        
    pdffitwindow=false,     
    pdfstartview={FitH},    
    pdftitle={My title},    
    pdfauthor={Author},     
    pdfsubject={Subject},   
    pdfcreator={Creator},   
    pdfproducer={Producer}, 
    pdfkeywords={keyword1} {key2} {key3}, 
    pdfnewwindow=true,      
    colorlinks=false,       
    linkcolor=red,          
    citecolor=green,        
    filecolor=magenta,      
    urlcolor=cyan           
}

\newenvironment{itemize*}
  {\begin{itemize}
    \setlength{\itemsep}{0pt}
    \setlength{\parskip}{0pt}}
  {\end{itemize}}

\newenvironment{enumerate*}
  {\begin{enumerate}
    \setlength{\itemsep}{0pt}
    \setlength{\parskip}{0pt}}
  {\end{enumerate}}

\newenvironment{description*}
  {\begin{description}
    \setlength{\itemsep}{0pt}
    \setlength{\parskip}{0pt}}
  {\end{description}}

\def\ben{\begin{enumerate*}}
\def\een{\end{enumerate*}}
\def\bi{\begin{itemize*}}
\def\ei{\end{itemize*}}
\def\bd{\begin{description*}}
\def\ed{\end{description*}}
\def\be{\begin{equation}}
\def\ee{\end{equation}}
\def\bea{\begin{eqnarray}}
\def\eea{\end{eqnarray}}
\def\bfl{\begin{flushleft}}
\def\efl{\end{flushleft}}

\begin{document}

\title{Was the Universe Actually Radiation Dominated Prior to Nucleosynthesis?}

\author{John T. Giblin, Jr.$^{1,2}$}
\email{giblinj@kenyon.edu} 
\author{Gordon Kane$^{3}$}
\email{gkane@umich.edu} 
\author{Eva Nesbit$^{4}$}
\email{ehnesbit@syr.edu} 
\author{Scott Watson$^{4}$}
\email{gswatson@syr.edu} 
\author{Yue Zhao$^{3}$}
\email{zhaoyhep@umich.edu}
\affiliation{$^{1}$~Department of Physics, Kenyon College, 201 N College Rd, Gambier, OH 43022}
\affiliation{$^{2}$~3CERCA/ISO, Department of Physics, Case Western Reserve University,
10900 Euclid Avenue, Cleveland, OH 44106}
\affiliation{$^{3}$~Michigan Center for Theoretical Physics, University of Michigan, Ann Arbor, MI 48109}
\affiliation{$^{4}$~Department of Physics, Syracuse University, Syracuse, NY 13244, USA}

\date{\today}

\begin{abstract}
Maybe not. String theory approaches to both beyond the Standard Model and Inflationary model building generically predict the existence of scalars (moduli) that are light compared to the scale of quantum gravity. These moduli become displaced from their low energy minima in the early universe and lead to a prolonged matter-dominated epoch prior to BBN. 
In this paper, we examine whether non-perturbative effects such as parametric resonance or tachyonic instabilities can shorten, or even eliminate, the moduli condensate and matter-dominated epoch. 
Such effects depend crucially on the strength of the couplings, and we find that unless the moduli become strongly coupled the matter-dominated epoch is unavoidable.
In particular, we find that in string and M-theory compactifications where the lightest moduli are near the TeV-scale that a matter-dominated epoch will persist until the time of Big Bang Nucleosynthesis.

\end{abstract}
\maketitle
\thispagestyle{empty}
Moduli are a generic prediction in string theoretic approaches to beyond the Standard Model \cite{Kane:2015jia} and inflationary model building \cite{Baumann:2014nda}.
It was noted long ago that these moduli could be displaced from their low-energy minima in the early universe, and their coherent oscillations lead to a period of matter domination
 \cite{Banks:1993en,deCarlos:1993wie,Coughlan:1983ci,Banks:1995dt,Banks:1995dp}.  This matter phase has important differences from a strictly thermal universe and is a rich source of dark matter phenomenology -- for a review see \cite{Kane:2015jia}.  The matter phase can also lead to enhanced growth of structure \cite{Erickcek:2011us,Fan:2014zua,Erickcek:2015bda}, changes in inflationary predictions for the cosmic microwave background \cite{Easther:2013nga}, and also the formation of 
primordial black holes \cite{Georg:2016yxa,Georg:2017mqk}.  
These cosmological and phenomenological predictions depend on the duration of the matter phase, which is determined by the moduli mass and couplings to other fields.  

It is expected that moduli couple gravitationally, and the matter phase will persist until the perturbative decay of the modulus completes which, for $50$ TeV moduli, will be around the  time of Big Bang Nucleosynthesis (BBN) \cite{Kane:2015jia}. In this paper, we want to revisit these assumptions and determine if effects such as parametric enhancement \cite{Kofman:1997yn,Traschen:1990sw} or tachyonic instabilities \cite{Felder:2000hj} can lead to an enhanced decay of the moduli. In the former case, as the field oscillates, particles are produced, and Bose-Einstein statistics can lead to a significant enhancement of the decay compared to the perturbative decay rate \cite{Traschen:1990sw,Kofman:1997yn} (for a review see \cite{Allahverdi:2010xz,Amin:2014eta}).  Whereas, in tachyonic resonance, if the mass squared of the field becomes negative due to the time and/or field dependence of the couplings this can lead to the efficient decay of the field in less than a single oscillation \cite{Felder:2000hj}.  It has also been argued that the dynamics and backreaction of the produced particles could be used to `trap' moduli \cite{Kofman:2004yc,Watson:2004aq,Greene:2007sa,Cremonini:2006sx}.  If these types of instabilities are present they can significantly enhance the moduli decay rate resulting in less of a matter phase or even prevent the formation of the moduli condensate all together.  For very light moduli -- that would decay after BBN -- this enhanced decay may lead to a new way to address the cosmological moduli problem \cite{Banks:1993en,deCarlos:1993wie,Coughlan:1983ci,Banks:1995dt,Banks:1995dp}.

\section{Moduli Decay through Parametric and Tachyonic Resonance}
The moduli will typically couple to other fields with gravitationally suppressed couplings.
This is the case in examples like KKLT \cite{Kachru:2003aw}, as well as the cases of Large Volume Compactifications in Type IIB \cite{Conlon:2005ki} 
and G2 compactifications of M-theory \cite{Acharya:2008zi}. The perturbative decay rate of the modulus is then $\Gamma  \sim m_\sigma^3/\Lambda^2$,
where $m_\sigma$ is the mass of the modulus and $\Lambda$ the suppression scale. Taking\footnote{We work with sign convention $(-,+,+,+)$ and with the reduced Planck mass $m_p=1/(8\pi G)^{1/2}=2.4 \times 10^{18}$ GeV. We use Greek indices to denote space-time $\mu=0,1,2,3$ whereas latin indices imply spatial directions only $k=1,2,3$.} $\Lambda \sim m_p$ the corresponding reheat temperature for a $m_\sigma = 50$ TeV scalar is 
around $5$ MeV \cite{Kane:2015jia}. Here we would like to determine whether parametric or tachyonic instabilities in the moduli can result in a faster decay and so higher reheat temperature.

We are motivated by recent work on preheating and the production of gauge fields at the end of inflation \cite{Deskins:2013lfx,Adshead:2015pva,Adshead:2016iae}.
In these papers it was found that a tachyonic instability to production of massless gauge fields from inflaton couplings $\sigma F_{\mu \nu} \tilde{F}^{\mu \nu} / \Lambda$ \cite{Adshead:2015pva,Adshead:2016iae} or 
$\sigma F_{\mu \nu} {F}^{\mu \nu} / \Lambda$ \cite{Deskins:2013lfx} can lead to explosive particle production and drain energy completely before the inflaton can complete a full oscillation. 
If this result were also true for moduli, then this could prevent the formation of the condensate and the matter-dominated phase.

\subsection{Moduli Coupling to Gauge Fields}
In all of the string constructions mentioned above there are moduli with masses generated by
gravitationally mediated Supersymmetry (SUSY) breaking.  The corresponding moduli mass is determined by the gravitino
mass $m_{3/2}$ as $m_\sigma = c \, m_{3/2}$ where $c$ is a constant determined by the particular string theory realization, e.g. in the G2 MSSM $c \simeq 2$. 

We now consider the coupling of the moduli to a hidden sector gauge field
\be
S=\int d^4x \sqrt{-g} \left( - \frac{1}{4} F_{\mu \nu} F^{\mu \nu} -  \frac{c}{4 \Lambda} \sigma \, F_{\mu \nu} F^{\mu \nu} \right), 
\label{action}
\ee
where $c$ is an order one constant (computable in a given string model) and consistency of the effective theory requires $\sigma < \Lambda$.
The corresponding equations of motion are
\bea \label{eqns}
\nabla_\mu F^{\mu \nu} + \frac{c}{\Lambda} \nabla_\mu \left( \sigma \, F^{\mu \nu} \right)=0, \\
\Box \sigma = \frac{\partial V}{\partial \sigma} + \frac{c}{4 \Lambda} F_{\mu \nu}F^{\mu \nu}.
\eea

Working in Coulomb gauge $A^0=0, \; \; \partial_i A^i=0$, neglecting the expansion of the background, and introducing 
the field redefinition $\tilde{A}_k=\left[ a(t) \left( 1+ c \, \sigma/\Lambda
 \right) \right]^{1/2}  A_k$ the resulting equations of motion are

\bea \label{phi_eom}
\ddot{\sigma}&+&3H \dot{\sigma} +m_\sigma^2 \sigma \nonumber \\
&=& \frac{c}{2 \Lambda} \left[ \frac{ \dot{A}_\mu \dot{A}^\mu}{a^2} + \epsilon_{\mu \nu \lambda} \epsilon_{\; \; \alpha \beta}^\lambda 
\nabla^\mu A^\nu \nabla^\alpha A^\beta \right]
\eea
\bea \label{Aeqn}
\ddot{\tilde{A}}_k &+& \left[ k^2 
+\frac{1}{2}\left( \frac{1}{1+c\, \sigma/\Lambda} \right)^2 \right. \nonumber \\
&\times&\left. \left( \frac{1}{2}c^2 \frac{\dot{\sigma}^2}{\Lambda^2} -c^2 \frac{\sigma \ddot{\sigma}}{\Lambda^2}-c \frac{\ddot{\sigma}}{\Lambda} \right) \right] \tilde{A}_k =0,
\eea
The moduli will remain frozen in their false minimum until $H \simeq m_{\sigma}$ at which time the moduli begin oscillations and
$\sigma(t)=\sigma_0 \cos(mt)$ where the initial amplitude is typically $\sigma_0 \sim m_p$.

The gauge field equation can be put in the form of a Mathieu equation by introducing the 
time variable $z=mt/2$.  Noting that consistency of the effective theory requires $\sigma_0 < \Lambda$ and keeping only the leading terms we have
\be \label{Azeqn}
\frac{d^2A_k}{dz^2}+\left[ 4 \left(\frac{k}{m_\sigma}\right)^2  + 2c \left( \frac{\sigma_0}{ \Lambda} \right)  \cos(2z) \right] A_k =0
\ee
where we have dropped terms further suppressed by powers of $\sigma_0/\Lambda$ and we note that the leading time-dependent mass term corresponds to the term $\sim \ddot{\sigma}/\Lambda$ in \eqref{Aeqn}.

Comparing \eqref{Azeqn} to the usual Mathieu equation
\be \label{m_eqn}
\frac{d^2u}{dz^2}+\left[ {\cal A}_k + 2q\cos(2z)\right]u=0,
\ee
suggests the identifications
\be \label{us}
{\cal A}_k\equiv 4 \left( \frac{k}{m_\sigma}\right)^2, \;\;\;\;\;\; q\equiv c \left(\frac{\sigma_0}{\Lambda}\right).
\ee
Tachyonic instability corresponds to the condition ${\cal A}_k < 2q$, broad resonance occurs for $q\gg1$ and narrow resonance occurs for $q\lesssim 1$.
We can immediately see that broad resonance is forbidden, since validity of the effective theory requires $\sigma_0<\Lambda$ or $q<1$. Moreover, although narrow resonance could play a role, it may not lead to significant enhancement of the production \cite{Kofman:1997yn}. Thus, we focus on the case of tachyonic resonance. 

\subsection{Tachyonic Resonance -- Analytic Treatment}
The modes that will undergo tachyonic resonance correspond to ${\cal A}_k < 2q$ in \eqref{m_eqn}, which for the identification \eqref{us} implies
\be \label{cond1}
k<\frac{1}{\sqrt{2}} \left( \frac{\sigma_0}{\Lambda} \right)^{1/2} m_\sigma.
\ee
However, for post-inflation we are interested in sub-Hubble modes\footnote{This is required by causality if the gauge modes begin in their vacuum state following inflation.} so we also require $k/H>1$ implying the modes 
of interest lie in a band
\be
1<\frac{k}{H}<\frac{1}{\sqrt{2}} \left( \frac{\sigma_0}{\Lambda} \right)^{1/2} \left( \frac{m_\sigma}{H} \right).
\ee
Thus, for tachyonic production of modes we require 
\be \label{estimate}
\frac{1}{\sqrt{2}} \left( \frac{\sigma_0}{\Lambda} \right)^{1/2} \left( \frac{m_\sigma}{H} \right) \gg 1,
\ee
so at the onset of the moduli phase, when $H \simeq m_\sigma$ perturbativity of the effective theory again seems to limit the level of enhancement in gauge field production,
since we require $\sigma_0 < \Lambda$. However, although the initial moduli displacement is typically expected to be an order of magnitude or so below the cutoff, as the moduli oscillations continue the Hubble parameter will continue to decrease $H<m_\sigma$, and tachyonic resonance becomes possible.  There is a competing effect that the amplitude of the moduli oscillations also decreases compared to its initial value $\sigma_0$.
It is a quantitative question of how important tachyonic resonance is for moduli decay and the duration of the epoch.  Moreover, during oscillations, creation of moduli (moduli particles, meaning $k\neq 0$ modes), particle scattering, and backreaction of both moduli and gauge fields can play an important role, as well as the expansion of the universe. To account for these complexities and non-linearities we perform a lattice treatment and present those results in the next section.

\subsection{Tachyonic Resonance -- Lattice Results}
To determine whether tachyonic (or parametric) instabilities occur in the system \eqref{phi_eom} and \eqref{Aeqn} we perform fully non-linear lattice simulations.  We build our simulations using the software {\sc GABE} \cite{Child:2013ria}, which has been used previously to study the interactions of scalar fields and U(1) Abelian gauge fields \cite{Deskins:2013lfx,Adshead:2015pva,Adshead:2016iae}.  
Our simulations allow us to account not only for gauge field production, but also the effects of scalar particle production, rescattering, backreaction, and the expansion of the universe. 

There are several restrictions on the allowed values of the fields and parameters of our model. For example, although we perform a lattice simulation, validity of the effective Supergravity description requires that the non-renormalizable operator in \eqref{action} remain subdominant to the leading kinetic term.  Since $c$ is a dimensionless  ${\cal{O}}(1)$ Wilson coefficient this requires that $\sigma$ not exceed the UV cutoff $\Lambda$ (which is typically order the Planck or string scale).  

We note that our simulations are similar to those of \cite{Deskins:2013lfx}, where the role of the inflaton there, is instead given by the moduli here.  As we will see, a key difference in our results compared to those of \cite{Deskins:2013lfx} is that there the authors considered a toy model with a dilatonic type coupling that could enter a ``strong coupling" regime.  In this paper, we are limited by the validity of the effective theory $\sigma < \Lambda$ and we'll see this limits our ability to establish a strong resonance behavior\footnote{The result that validity of an effective field theory approach can limit the importance of parametric resonance was noted recently in \cite{Giblin:2017qjp}.}.

In order to establish as large a resonance as possible we will take the initial amplitude of the moduli to be near the Planck scale $\sigma_0 \simeq m_p$ (we take $\sigma_0 = 0.2\,m_{\rm pl}$ as a fiducial value).
Then, given our discussion of the validity of the effective theory requires that we take $\Lambda \sim m_p$, and as the field can change sign this also ensures that the 
kinetic term of \eqref{action} retains the correct sign. 
This limits us to a maximum coupling $c/4\Lambda \approx 6.9\,m_{\rm pl}^{-1}$.  Throughout this section we will use this maximum value as to make the potential tachyonic window as large as possible (we have checked that for lower values of the cutoff the resonance is even weaker than the results we present here).  We are left with only one free parameter, $m_\sigma$, which also sets the Hubble scale at the beginning of coherent oscillations.

Using {\sc GABE} we discretize space onto a grid of $128^3$ points that are on a homogeneously expanding box.  The box has initial size, $L = 4 m_\sigma^{-1} \approx 2 H_0^{-1}$.  The simulations solve \eqref{phi_eom} and \eqref{Aeqn} along with the Friedmann equations. 
For numerical simplicity, we employ the standard {\sl unit-less conformal} time, $d\tau = a(t) \,m\,dt$. We use an adaptive time step, $\Delta \tau = 0.005/a(\tau)$ so that we resolve the co-moving modes throughout the simulation.  We initialize the modulus field consistent with the expectations of a field that carries the ``freeze out'' power as modes re-enter the horizon\footnote{We start our simulations at the beginning of moduli oscillations and we take adiabatic initial conditions so that the inflaton fluctuations will have been transferred to the moduli that come to dominate the energy density (we assume no isocurvature, however see \cite{Iliesiu:2013rqa}) and assume that $\Delta_s^2\approx 10^{-10}$.},
\be
\left<\delta \sigma (k)\delta\sigma (k^\prime)\right> =\frac{\pi^2}{2}\left( \frac{\Delta_s^2\sigma_0^2 }{H_0^3}\right)  \, \delta\left( k-k^\prime\right),
\ee
assuming that most modes have not grown much since horizon re-entry\footnote{Prior to moduli domination we take the universe to be radiation dominated following inflationary reheating and sub-Hubble modes of the moduli will undergo very little growth (their perturbations grow logarithmically with the scale factor $\sim \log(a)$.} and have recently re-entered ($k\approx H_0$).  
For the gauge fields we set the initial conditions consistent with the Bunch Davies vacuum \cite{Deskins:2013lfx},
\be
\left<\left | A_i(k)A_j(k^\prime)\right|^2\right>  =  \frac{\delta_{ij}\, \delta\left( k-k^\prime\right)}{2a\left( 1+ c\sigma/\Lambda \right)},
\ee
with zero homogenous mode (we comment on the robustness of this assumption shortly).  
We take the initial surface in Coulomb gauge, but the rest of the simulation is carried out in Lorenz gauge, $\partial^\mu A_\mu = 0$, where Gauss' constraint is treated as a dynamical degree of freedom (as the equation of motion for $A_0$) and we check that the gauge constraint is maintained throughout our simulations.  As we increase the mass of the modulus field, we shrink the physical size of the Hubble patch at the beginning of the simulation.  This is the best approach to resolving shorter wavelength modes of the gauge fields, and hence, a larger fraction of energy in the gauge sector.  As we set the initial conditions, we impose a window function (as in \cite{Deskins:2013lfx}) that cuts off power to modes $k \gtrsim 90\,m_\sigma$ for numerical stability. However, this scale is above the scale at which we would expect to see tachyonic instabilities.

\begin{figure}[htbp]\
\centering
\includegraphics[width=\columnwidth]{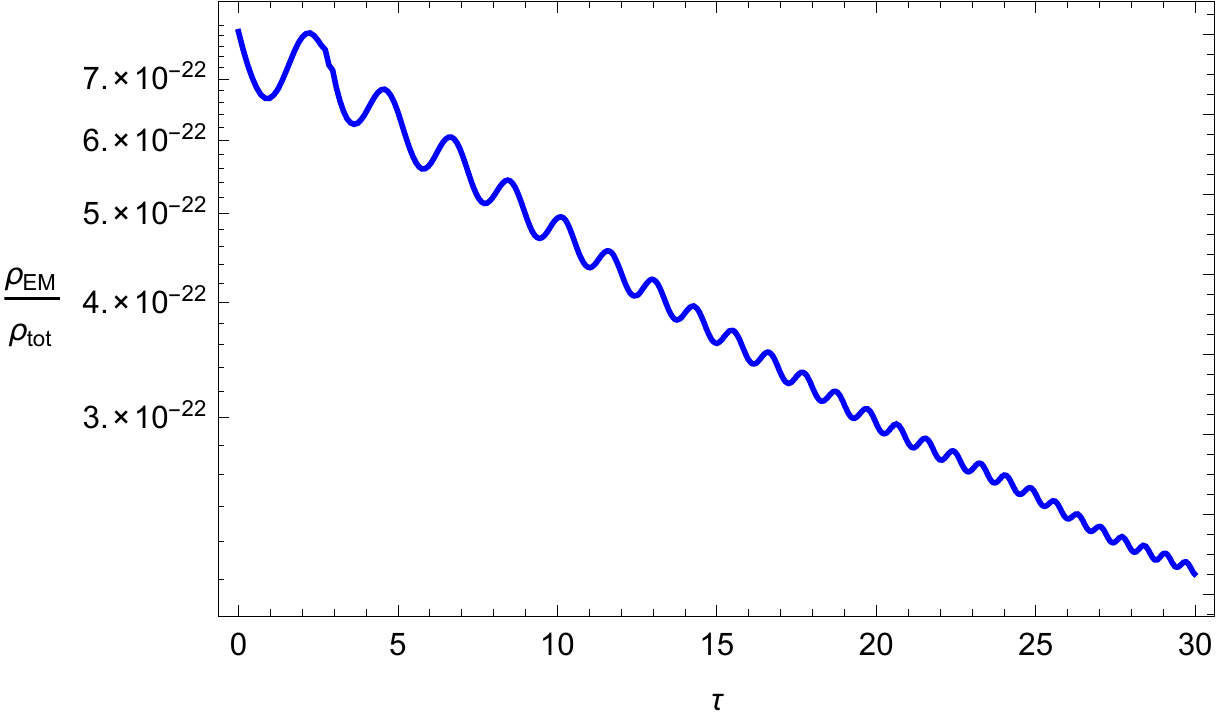}
\includegraphics[width=\columnwidth]{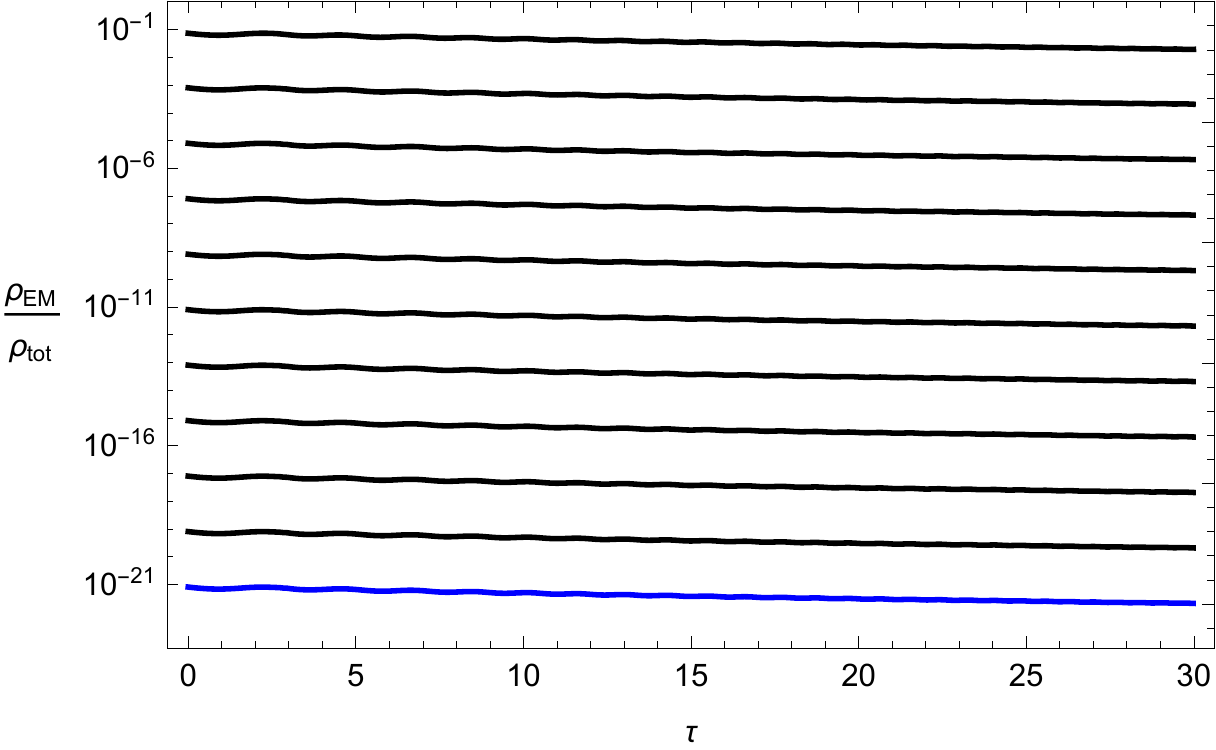}
\caption{Plot of $\rho_{\rm EM}/\rho_{tot}$ vs. unit-less conformal time (see text) for a set of maximally coupled simulations, $c/\Lambda = 6.7\,m_{p}$.  The top panel shows a simulation of the fiducial value of $m_\sigma  = 50\,{\rm TeV}$ and the bottom panel shows a range of masses, from $m_{\sigma}=50\,{\rm TeV}$ (bottom) to $m_{\sigma}=5\times 10^{11} \,{\rm TeV}$, the $50\,{\rm TeV}$ case is labeled in blue in both plots.  For each simulation $\rho_{tot}(t)$ is approximately the same, since the energy of the modulus is dominated by its homogeneous mode and is always the dominant component.}
\label{fig:moneyplot1}
\end{figure}

Following \cite{Deskins:2013lfx}, we take the ratio of the gauge field energy density ($\rho_{\rm EM}$) to the total energy  ($\rho_{\rm tot}$) as a figure of merit of the amplification of the gauge field and the effectiveness of the tachyonic (and parametric) instabilities.  Figure \ref{fig:moneyplot1} shows the evolution of this parameter as a function of time for a large range of moduli masses. 
We find the robust result that {\sl regardless} of the (relative) amplitude of the initial fluctuations of the gauge fields, tachyonic (and parametric) instabilities are absent and do not lead to significant amplification of the gauge fields.  The variation in the initial value of $\rho_{\rm EM}$ reflects that we allow for different values of the moduli mass as discussed above.  Considering a pre-existing density of gauge modes ({\it e.g.} non-Bunch Davies initial conditions with modes that were classically or quantum mechanically excited during inflation\footnote{Model independent bounds on the level of gauge field production during inflation was recently established in \cite{Green:2015fss}.  There it was shown that requiring successful inflation limits the amplification of gauge fields which here limits the size of the initial amplitude taken for the gauge fields, {\it i.e.} one can not take the initial amplitude to be arbitrarily large.}) would have a similar effect, amplifying the initial spectrum of the gauge field, and hence, raising $\rho_{\rm EM}/\rho_{\rm tot}$ on the initial surface. 

\begin{figure}[htbp]\
\centering
\includegraphics[width=\columnwidth]{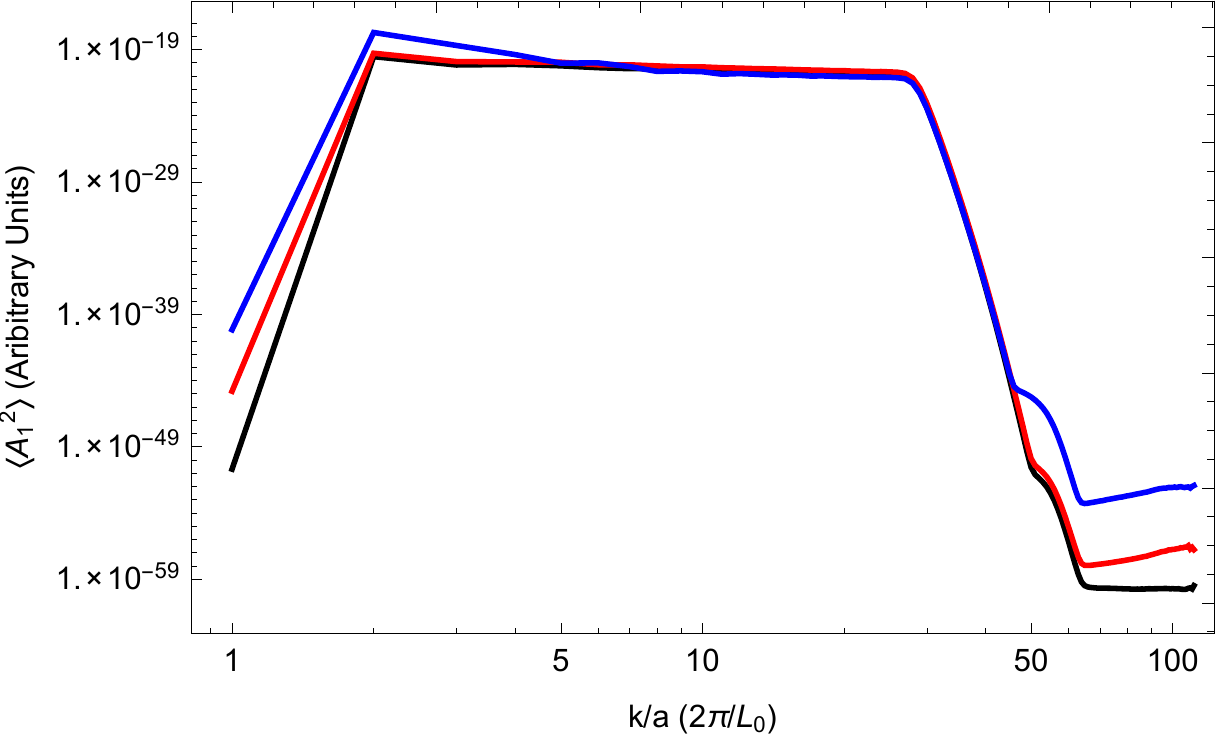}
\caption{The power spectra of one component of the gauge field, $A_1$ at the beginning of the simulation (black), at the first zero crossing (red) and at the second zero crossing (blue) in a simulation where  $m_\sigma = 50\,{\rm TeV}$.  At higher frequencies, the power is suppressed due to the window function imposed on the initial slice, the slight increases in these frequencies is not a physical response, but an accumulation of numerical truncation errors (and is still many orders of magnitude below the scales of interest).  The increase in the zero-momentum bin is a consequence of the initial value being set to zero to machine-precision, with truncation errors making it drift away.  The spectra undergo negligible amplification over the course of the simulation.  The other spatial components of the field have identical behaviors, and similar results are seen in all simulations.  We find no indication of tachyonic or parametric instabilities.}
\label{fig:specz}
\end{figure}

An additional measure at which to look for instabilities is in the spectra of the coupled fields.  In Figure~\ref{fig:specz}, we see that there is very little change to the power spectra of the fields.  In cases where instabilities exist, we can generally see these instabilities in the power spectra of the fields.  In none of the cases we studied did we see any indication of tachyonic or parametric instabilities.

\begin{figure}[htbp]\
\centering
\includegraphics[width=\columnwidth]{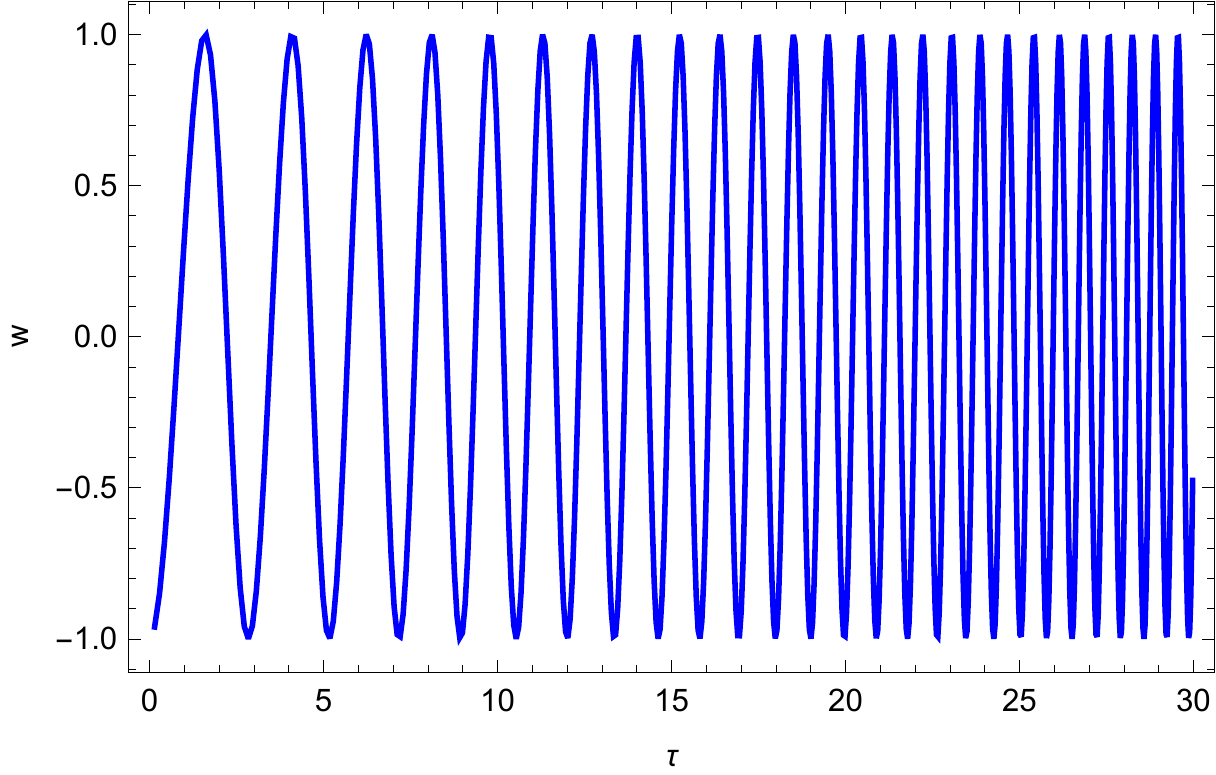}
\caption{The equation of state for a simulation where  $m_\sigma = 50\,{\rm TeV}$ vs. unit-less conformal time (see text).
We see that the average of the equation of state is that of a matter-dominated universe. }
\label{fig:eos}
\end{figure}

Although we have not found significant evidence for an increased decay of the moduli, this does not necessarily imply a matter-dominated epoch.
Indeed, it was recently shown that the non-linear dynamics of the fields can have an important influence on the equation of state \cite{Lozanov:2016hid}.
Thus, we must lastly ensure that the expansion mimics that of a matter-dominated single-component universe.  To do this, we track the equation of state parameter, $w=p/\rho$, which is the usual ratio of the isotropic pressure to the energy density.  Figure~\ref{fig:eos} shows this for the fiducial case, $m_\sigma = 50\,{\rm TeV}$, and shows that $w$ oscillates, as expected, between $\pm1$ as is the case of a massive scalar field dominated by its homogeneous value.  

\section{Comments and Conclusions}
In this paper, we have considered the coupling of moduli to hidden sector gauge fields for a range of masses and initial values of the gauge fields.
We found that even as we approach modestly strong coupling, tachyonic and parametric instabilities have no effect on the moduli decay rate.
Moreover, we have seen that the equation of state during the moduli oscillations averages to the previously anticipated result of a matter-dominated universe.
As gauge field production relies on the moduli dynamics breaking the conformal invariance of the gauge field sector \cite{Demozzi:2009fu}, and in these string motivated models the source of this breaking comes from non-renormalizable operators, it may not be that surprising that this effect turned out to be negligible.  
One reason for considering these operators was that such couplings generically appear in string theories, and are model independent in the sense that
they arise strictly in the moduli sector and are typically independent of how one embeds the visible sector.
This is indeed the case in examples like KKLT \cite{Kachru:2003aw}, as well as the cases of Large Volume Compactifications in Type IIB \cite{Conlon:2005ki} 
and G2 compactifications of M-theory \cite{Acharya:2008zi}.

One may wonder if more model dependent couplings (arising from embedding the visible sector in a particular string construction) may alter our conclusions.
For example, moduli couplings to the Higgs ($\sim\sigma H^\dagger H$) are relevant operators and the moduli might undergo enhanced decay to Higgs bosons.
However, such couplings were already considered some time ago by Brandenberger and Shuhmaher in \cite{Shuhmaher:2007pv,Shuhmaher:2005mf}.  
They considered relevant operators arising from SUSY breaking for a range of moduli masses.  
Their results are similar to our findings for non-renormalizable operators.
That is, if one requires a perturbative theory and consistency of the effective field theory then both parametric and tachyonic resonance 
does not significantly alter the moduli decay rate.  

Our results, as well as those of  \cite{Shuhmaher:2005mf}, suggest that if one is to eliminate the moduli dominated
epoch one is going to have to consider moduli that are strongly coupled.
There is some motivation for this in string theory \cite{Banks:1994sg} (for more recent work see \cite{DelZotto:2016fju}), however there must typically be at least one light modulus
if we are to realize the perturbative Standard Model in a string construction \cite{Cremonini:2006sx}.
For this reason, we take our results as a robust prediction that string theories lead to the expectation for a prolonged, matter-dominated epoch prior to BBN.

\section*{Acknowledgements}
We are grateful to Peter Adshead, Bhaskar Dutta, Adrienne Erickcek and Matt Reece for useful discussions.  
J.T.G. is supported by the National Science Foundation, PHY-1414479.
S.W. thanks the Michigan Center for Theoretical Physics for hospitality.
S.W. is supported in part by NASA Astrophysics Theory Grant NNH12ZDA001N and DOE grant DE-FG02-85ER40237.
Y.Z. is supported by DOE grant DE- SC0007859.
This work was completed at the Aspen Center for Physics, which is supported by National Science Foundation grant PHY-1066293.

\bibliographystyle{apsrev4-1}

%

\end{document}